\documentclass{article}
\usepackage{iclr2027_conference,times}

\usepackage{amsmath,amsfonts,bm}

\def\eqref#1{equation~\ref{#1}}

\def\1{\bm{1}}

\DeclareMathAlphabet{\mathsfit}{\encodingdefault}{\sfdefault}{m}{sl}
\SetMathAlphabet{\mathsfit}{bold}{\encodingdefault}{\sfdefault}{bx}{n}

\usepackage[T1]{fontenc}
\usepackage[utf8]{inputenc}
\usepackage{textcomp}
\usepackage{hyperref}

\usepackage{booktabs,graphicx}
\usepackage{url}
\usepackage{fancyvrb}
\usepackage{longtable,booktabs,array,calc}
\usepackage{needspace}

\hypersetup{
    hidelinks,
    pdftitle={ContextTom: Training-Free Video VLM Acceleration with Early Spatial Context},
    pdfauthor={Yuqi Li, Neil Rodrigues, Xianglong Wang, Junming Liu, Yi Xiao, Weilun Feng, Chuangguang Yang, Yingli Tian, Zhongwei Cheng, Wei Li, Lin Chen}
}

\providecommand{\tightlist}{%
  \setlength{\itemsep}{0pt}%
  \setlength{\parskip}{0pt}%
}

\title{
ContextTom: Training-Free Video VLM Acceleration\\
with Early Spatial Context
}

\author{
Yuqi Li$^{1,2}$,
Neil Rodrigues$^{1}$,
Xianglong Wang$^{1}$,
Junming Liu$^{3}$,
Yi Xiao$^{4}$,
Weilun Feng$^{5}$, \\
Chuangguang Yang$^{5}$,
Yingli Tian$^{2}$,
Zhongwei Cheng$^{1}$,
Wei Li$^{1}$,
Lin Chen$^{1}$ \\
\\
$^{1}$Wyze Labs, Inc. \\
$^{2}$The City College of New York, CUNY \\
$^{3}$Tongji University \\
$^{4}$University of Alabama at Birmingham \\
$^{5}$Institute of Computing Technology, Chinese Academy of Science
}

\iclrfinalcopy

\begin{document}

\maketitle

\lhead{Preprint}

\begin{abstract}
Video vision-language models process thousands of visual tokens before producing an answer. Reducing these tokens early saves computation, but also determines which evidence reaches the language model. We introduce \textbf{ContextTom}, a training-free framework that separates the timing of temporal and spatial compression. Its central idea is to compress time before vision encoding while preserving spatial coverage through early multimodal processing. A fixed-budget temporal partition produces representative medoid anchors, supplemented with bounded local detail from neighboring observations. The anchors retain their complete spatial grids through a short decoder prefix, allowing question tokens to access the visual evidence and completing Qwen3-VL\textquotesingle s multi-level visual injections. ContextTom then selects spatial representatives using query attention and agreement across visual feature levels, and continues directly from their computed states. This schedule reduces vision computation and later decoder workload while reusing the prefix. On Qwen3-VL-8B, ContextTom achieves 76.22\% accuracy on
Video-MME short with a mean time to first token (TTFT) of
123.5\,ms, yielding a $2.77\times$ speedup over full-input
inference.
It matches the full-input accuracy of 72.20\% on EgoSchema
with a $2.82\times$ speedup and achieves a $5.54\times$
speedup on LongVideoBench at 59.24\% accuracy.

\end{abstract}

\hypertarget{introduction}{%
\section{Introduction}\label{introduction}}

Video vision-language models (VLMs) support temporal reasoning
and question answering~\citep{tang2025video_survey}, but long
visual sequences make vision encoding and language-model
prefill costly~\citep{jin2025efficient_survey}.
Training-free token compression reduces this workload while
keeping pretrained weights fixed~\citep{shao2026token_survey}.
The key design decision is not only how many tokens survive,
but \textbf{how much computation they receive before removal.}

Existing methods expose complementary opportunities. ToMe reduces visual redundancy through token merging; VisionZip compresses visual features before language processing; and FastV uses early decoder attention to identify tokens for later layers \citep{tome,visionzip,fastv}. Video methods additionally exploit repetition across frames. In particular, EarlyTom brings temporal compression into the vision encoder, saving computation before the language model \citep{earlytom}. Together, these approaches motivate treating the compression boundary as part of the algorithm.

This boundary becomes especially consequential in Qwen3-VL. Its DeepStack mechanism injects features from several vision-encoder depths into early decoder layers \citep{qwen3vl}. Spatial pruning before these layers limits both the locations receiving visual injections and the evidence accessible to question tokens. Retaining every token throughout the decoder, however, carries this cost into all later layers. The opportunity is to preserve spatial access during early multimodal processing, then reduce the sequence once that computation can guide selection.

We develop this idea through a division of labor between time and space. Temporal compression constructs a compact set of observations before expensive visual processing. Spatial compression selects which locations continue after a short decoder prefix. The temporal stage separates change detection from representative selection: change-and-drift segmentation allocates a fixed number of intervals, and a medoid represents each interval. Nearby observations supplement a small number of anchor cells with bounded residual features. This provides a representative temporal input while retaining selected local detail.

The spatial stage keeps each anchor\textquotesingle s entire grid through the first three decoder blocks, completing the model\textquotesingle s DeepStack injections. Question tokens can attend to these grids before selection. We then combine query importance with local similarity across final and intermediate visual features. A video-wide removal budget allocates compression across frames, while a per-frame floor preserves distributed token coverage. The surviving decoder states continue directly into later layers, together with their original positions and the caches already computed for the prefix.

We call this framework \textbf{ContextTom}. It coordinates three decisions: which temporal observations reach the vision blocks, when spatial selection takes place, and which computed states continue. The schedule targets both major inference costs. Temporal anchoring reduces the number of grids processed by
every vision-transformer block, while spatial selection
shortens the sequence processed by later decoder layers.
Direct continuation from the retained prefix states reuses
the computation that guides selection.

On the 900-question short subset of Video-MME~\citep{videomme},
ContextTom with Qwen3-VL-8B achieves 76.22\% accuracy and
123.5\,ms mean time to first token (TTFT), corresponding to
a $2.77\times$ speedup over full-input inference.
Across the short, medium, and long subsets, the relative
speedup remains close to $2.8\times$, while absolute latency
savings increase with video duration.
Compression-timing analysis shows that preserving the three-block
full-spatial prefix improves accuracy by 0.60 percentage points over
pre-decoder spatial compression for an additional 6.1\,ms TTFT.
Temporal-budget experiments reveal diminishing accuracy returns from
further retention, motivating a compact set of informative anchors.

Our contributions are threefold:

\begin{enumerate}
\def\labelenumi{\arabic{enumi}.}
\tightlist
\item
  \textbf{Representative temporal compression.} Fixed-budget segmentation, medoid anchors, and bounded local supplementation reduce the input to all vision-transformer blocks.
\item
  \textbf{Spatial selection after early visual injection.} Complete anchor grids remain available through the DeepStack prefix; query importance and multi-level similarity then select spatial representatives with global allocation and local coverage.
\item
  \textbf{Continuation from computed states.} A single decoder traversal reuses prefix states, original positions, and layer-specific KV histories, connecting the compression schedule to an efficient implementation.
\end{enumerate}

\hypertarget{related-work}{%
\section{Related Work}\label{related-work}}

\textbf{Visual token compression.}
ToMe merges similar tokens within vision transformers, while VisionZip
selects and consolidates visual features before language-model processing
\citep{tome,visionzip}.
FastV uses early decoder attention to prune visual tokens for subsequent
layers \citep{fastv}.
AIM combines iterative similarity-based token merging before the language
model with progressive pruning based on multimodal importance within
its layers \citep{aim}.
PriorTR separates task-conditioned information from
model-induced attention priors for visual token
ranking~\citep{chen2026priortr}.
ContextTom combines early temporal reduction with spatial selection after
multi-level visual injection, retaining the computed states of spatial
representatives.

\textbf{Video redundancy.}
Swift Sampling selects informative frames using Taylor prediction
residuals along visual feature trajectories \citep{swiftsampling}.
KTV performs question-agnostic clustering of frame features to select
keyframes, followed by visual token selection based on importance and
redundancy \citep{ktv}.
FastVID uses temporal segmentation and density-based pruning; HoliTom
combines temporal and spatial reduction outside the language model
with merging inside it \citep{fastvid,holitom}.
FrameFusion combines similarity-based merging and importance-based
pruning across decoder depths \citep{framefusion}.
VidCom$^{2}$ adjusts compression intensity across frames according to
their visual uniqueness \citep{vidcom2}.
ContextTom selects temporal anchors using inexpensive pixel descriptors
before the vision-transformer blocks, then uses query importance and
multi-level visual similarity to select spatial representatives after
the decoder's visual-injection prefix.
A shared spatial budget lets frames contribute different numbers of
surviving tokens.

\textbf{Early compression and visual injection.} DeepStack distributes visual tokens across language-model
layers~\citep{meng2024deepstack}. EarlyTom is the closest starting point. It merges frames within the vision encoder and applies spatial selection policies to dynamic and static frames \citep{earlytom}. ContextTom uses fixed-budget medoid anchors before the first vision block and moves spatial selection to the end of the DeepStack injection prefix in Qwen3-VL \citep{qwen3vl}. This design coordinates the compression boundary with multi-level local similarity, per-frame coverage, and direct continuation from selected decoder states.

\hypertarget{method}{%
\section{Method}\label{method}}

\begin{figure}[t]
    \centering
    \includegraphics[width=\linewidth]{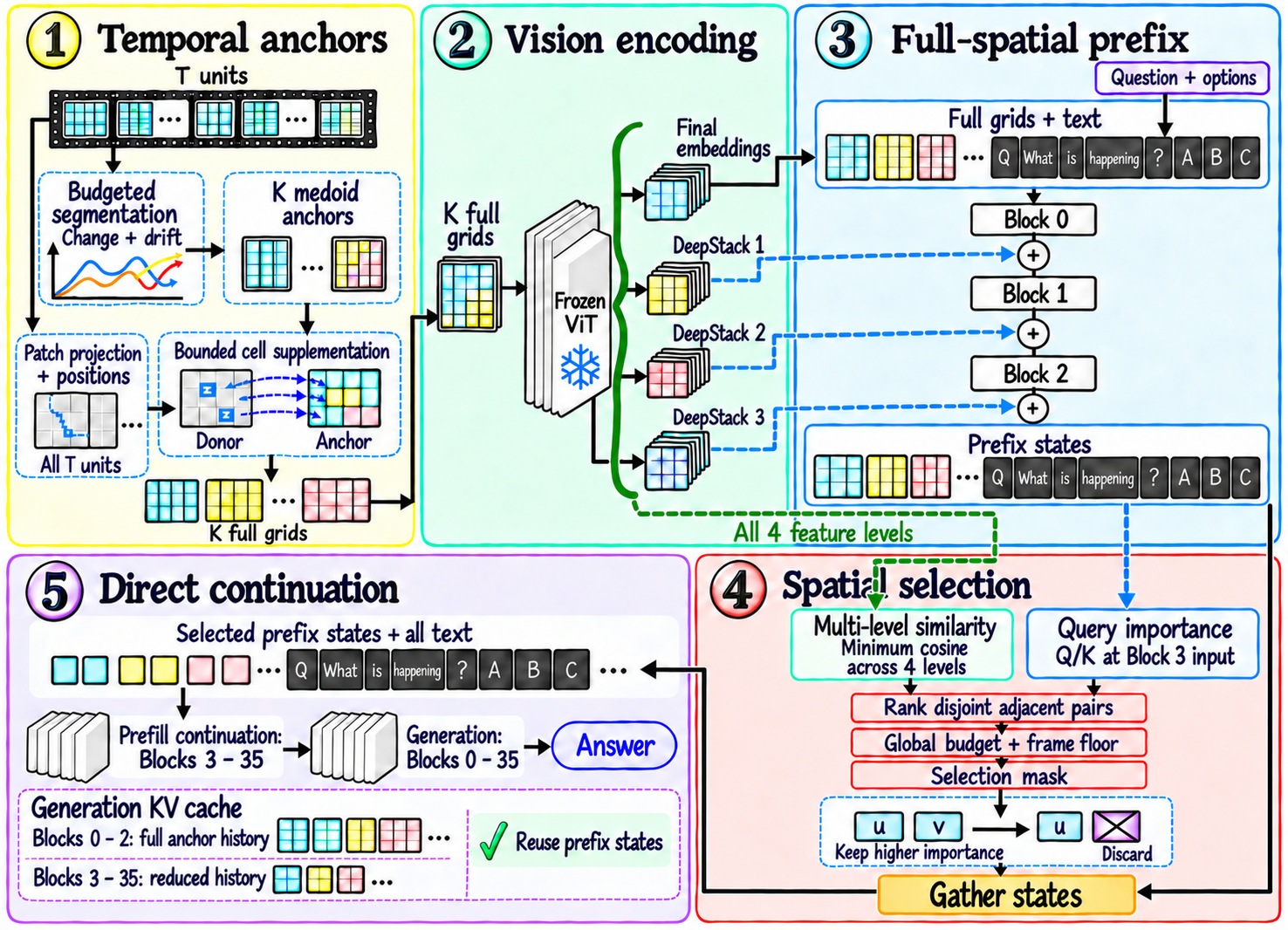}
    \caption{
    \textbf{Overview of the ContextTom inference pipeline.}
    Temporal medoid anchors receive bounded local supplementation
    before vision encoding. Their complete spatial grids, together
    with all text tokens, pass through a three-block decoder prefix
    with DeepStack injections. Query importance and multi-level
    visual similarity then guide spatial representative selection
    under a video-wide budget and per-frame retention floors.
    Selected prefix states continue directly through the remaining
    decoder blocks, preserving their original positions and all
    nonvisual tokens. Generation maintains full early-layer and
    reduced later-layer KV histories.
    }
    \label{fig:pipeline}
\end{figure}

\hypertarget{overview-compress-time-early-preserve-space-through-the-prefix}{%
\subsection{Overview: compress time early, preserve space through the prefix}\label{overview-compress-time-early-preserve-space-through-the-prefix}}

Consider a processed video with \(T\) temporal units and \(S=HW\) visual cells per unit. A temporal unit is a processor-level temporal patch, typically containing two sampled RGB frames in Qwen3-VL. Each visual cell corresponds to the \(m\times m\) patch group consumed by the spatial merger, with \(m=2\). Let \(M\) count all nonvisual prompt tokens, and let \(L\) be the decoder depth.

ContextTom retains \(K=\max\{1,\min(T,\operatorname{round}(r_tT))\}\) temporal anchors. The vision blocks encode their full spatial grids. The first \(p\) decoder blocks then process \(n_0=M+KS\) tokens; spatial selection reduces the remaining blocks to \(n_1=M+N_s\). All nonvisual tokens are retained. Our Qwen3-VL-8B configuration uses \(L=36\), \(p=3\), temporal retention \(r_t=0.52\), spatial target \(r_s=0.85\), and per-frame retention floor \(f=0.80\).

\textbf{Pipeline.}
As illustrated in Figure~\ref{fig:pipeline}, ContextTom performs
temporal segmentation and anchoring, reduced vision encoding,
full-spatial DeepStack prefix processing, query-guided spatial
selection, and continuation from retained states.
All model weights remain fixed.

\hypertarget{temporal-anchors-with-local-detail}{%
\subsection{Temporal anchors with local detail}\label{temporal-anchors-with-local-detail}}

\textbf{Allocate intervals, then choose representatives.} We obtain an inexpensive descriptor \(x_t\) by pooling per-cell RGB means to at most an \(8\times8\) map. Let \(d\) be mean absolute descriptor distance. Starting from \([0,T)\), we greedily split intervals using a score that combines adjacent change and drift from the current interval\textquotesingle s start:

\[
b(t\mid a,b)=d(x_t,x_{t-1})+\lambda_d d(x_t,x_a),
\qquad a<t<b.
\tag{1}
\]

At each step, the globally highest-scoring candidate is split, and scores are updated in its children. Splitting stops at exactly \(K\) intervals. Adjacent change captures abrupt transitions; start-relative drift gives gradual changes a role in budget allocation. Within each interval \(I_j\), the anchor is its medoid, \(a_j=\arg\min_{t\in I_j}\sum_{u\in I_j}d(x_t,x_u)\). This separates where an interval begins from which observation represents it.

\textbf{Supplement selected anchor cells.} Nearby units within the same interval can contribute detail at corresponding spatial cells. For each anchor cell, we select the donor with the largest product of relative detail gain and a bounded novelty factor. Only the top \(\lfloor\rho S\rfloor\) cells receive a contribution. Their patch features are updated as

\[
\widetilde z_{j,c}
=z_{a_j,c}+\beta_{j,c}\big(z_{u^*(c),c}-z_{a_j,c}\big),
\qquad 0\le\beta_{j,c}\le\eta/e.
\tag{2}
\]

Here \(u^*(c)\) is the selected donor, and unselected cells use \(\beta_{j,c}=0\). Features \(z\) include patch projection and visual position embeddings. The bounded coefficient keeps the anchor dominant; the cell budget concentrates supplementation on selected locations. We use radius \(R=2\), cell fraction \(\rho=0.25\), strength \(\eta=0.25\), and drift weight \(\lambda_d=1\). Appendix A specifies the scoring and tie-breaking rules.

Fusion occurs before the first vision-transformer block. Consequently, all vision blocks process \(K\) grids, while patch projection processes the original \(T\) units. Final visual embeddings \(V^{(0)}\) and DeepStack features \(V^{(1)},\ldots,V^{(J)}\) are produced from the same compressed stream, maintaining alignment across feature levels.

\hypertarget{preserve-spatial-access-through-visual-injection}{%
\subsection{Preserve spatial access through visual injection}\label{preserve-spatial-access-through-visual-injection}}

We assemble the prompt using all \(S\) cells of every anchor, the original text and timestamps, and the original multimodal position IDs \(\Pi\). The first \(p\) decoder blocks process this sequence with the model\textquotesingle s usual DeepStack additions. In Qwen3-VL-8B, \(J=p=3\), so spatial selection follows completion of all three visual injections.

This prefix gives question tokens access to the complete spatial grids of the retained temporal observations. It also applies every DeepStack injection before spatial reduction. The resulting hidden states \(H^{(p)}\) serve two purposes: they supply query importance for selection and become the starting states for the remaining decoder blocks.

At the input to block \(p\), we use its pretrained normalization, Q/K projections, head normalization, and rotary positions to score the text suffix following the last video-end marker. Attention is normalized over all causally visible keys. Averaging over suffix queries and heads, followed by normalization across visual positions, gives each cell an importance score \(s_i\). This suffix includes the question, answer options, and trailing instructions. Computing only its query rows supplies selection scores without another full decoder pass.

Causal attention lets question states incorporate visual evidence; query relevance enters the spatial decision through \(s_i\). The last DeepStack addition is present at the scoring boundary. Appendix B gives the exact attention and injection definitions.

\hypertarget{select-spatial-representatives-under-coverage-constraints}{%
\subsection{Select spatial representatives under coverage constraints}\label{select-spatial-representatives-under-coverage-constraints}}

\textbf{Combine local redundancy and query importance.} Within each anchor, we form disjoint adjacent cell pairs along the longer grid dimension. Unpaired boundary cells remain. For each pair \((u,v)\), we combine the minimum cosine similarity across visual feature levels with the normalized ascending ranks \(r_u,r_v\in[0,1]\) of query importance:

\[
\gamma_{uv}
=\min_{j\in\{0,\ldots,J\}}
  \cos\big(V^{(j)}_u,V^{(j)}_v\big)
-\lambda_q\min(r_u,r_v).
\tag{3}
\]

Larger \(\gamma_{uv}\) makes a pair more attractive to reduce. The minimum across levels favors pairs whose similarity persists across the visual hierarchy. Penalizing the smaller importance rank protects pairs in which both members matter to the query. A pair with one important member can still be reduced by retaining that member. We use \(\lambda_q=0.7\).

\textbf{Allocate a video-wide budget.} With \(P\) candidate pairs per frame, the per-frame removal capacity is \(C=\min\{P,S-\lceil fS\rceil\}\). The total removal budget is \(B=\min\{KC,KS-\lceil r_sKS\rceil\}\). We keep the top \(C\) candidates in each frame and select the global top \(B\) from their union. This allocates more removals to frames with stronger redundancy scores while retaining at least \(\lceil fS\rceil\) cells in every anchor.

Each selected pair retains the cell with higher query importance, yielding \(N_s=KS-B\) survivors. The corresponding state is gathered directly from \(H^{(p)}\). Spatial selection therefore preserves a computed representative state, while temporal supplementation constructs the input to the vision blocks. Pair construction, stable ranks, and tie-breaking are detailed in Appendix C.

\hypertarget{continue-from-the-retained-states}{%
\subsection{Continue from the retained states}\label{continue-from-the-retained-states}}

Let \(\mathcal S\) contain all nonvisual positions and surviving visual positions in sequence order. We gather hidden states and position IDs together and continue:

\[
\widetilde H^{(p)}=H^{(p)}_{\mathcal S},
\qquad \widetilde\Pi=\Pi_{\mathcal S},
\qquad
\widetilde H^{(\ell+1)}
=\mathcal L_\ell(\widetilde H^{(\ell)};\widetilde\Pi),
\quad p\le\ell<L.
\tag{4}
\]

This directly reuses the prefix computation. Retained cells keep their temporal and spatial coordinates, and every text token remains in the prompt. The first \(p\) layers keep their \(n_0\)-token KV histories; later layers cache the reduced \(n_1\)-token sequence. During generation, each layer appends the new token to its own history. Appendix C specifies positional bookkeeping and the complete inference algorithm.

The schedule reduces the number of grids processed by every vision block and the sequence length seen by \(L-p\) decoder blocks. Appendix~\ref{sec:development-token-flow} records visual-token counts at both compression boundaries.

The implementation combines this schedule with batched vision attention, fused normalization and rotary operations, and compact sequence assembly. It targets single-request SDPA inference. Model-side TTFT includes selection overhead and the complete path to first-token logits, measured after video decoding and preprocessing. Appendix D gives the cost decomposition, implementation settings, and timing protocol.


\section{Experiments}
\label{sec:experiments}

\paragraph{Benchmarks and Models.}
We evaluate on Video-MME~\citep{videomme}, EgoSchema~\citep{egoschema},
LongVideoBench~\citep{longvideobench}, and MVBench~\citep{mvbench}.
The benchmark tables cover Qwen3-VL-8B and Qwen3-VL-2B~\citep{qwen3vl},
InternVL3.5-8B~\citep{internvl35}, and
LLaVA-OneVision-7B~\citep{llavaonevision}.
The evaluation sets contain 2,700 Video-MME questions
(900 per duration subset), 500 labelled EgoSchema questions,
and 1,337 LongVideoBench validation questions.
MVBench runs include 3,600 or 3,800 questions according to media availability;
Tables~\ref{tab:contexttom_main} and~\ref{tab:contexttom_fixedframes}
identify the corresponding sample counts.

\paragraph{Evaluation Protocol.}
Qwen models use 1 FPS sampling; InternVL and LLaVA use 32 frames per video.
All experiments run on NVIDIA H100 GPUs. We report accuracy (\%),
mean time to first token (TTFT, ms), and speedup relative to full-input TTFT.
Component, compression-timing, and budget analyses use Qwen3-VL-8B
on Video-MME short. The default configuration is
$r_t=0.52$, $r_s=0.85$, $f=0.80$, and $p=3$.
Development profiling uses single-request eager SDPA inference, 1 FPS,
a 512-pixel long edge, and no subtitles. An untimed same-input warmup
precedes each fresh synchronized forward pass.
Model-side TTFT includes selection, vision encoding, sequence assembly,
and decoder processing through first-token logits; end-to-end latency
also includes preprocessing and answer generation.
Appendix~\ref{execution-and-measurement} gives profiling and metric details.

\paragraph{Research Questions.}
We examine \textbf{RQ1}, the accuracy--latency trade-off across video benchmarks;
\textbf{RQ2}, the processing stages responsible for efficiency gains;
\textbf{RQ3}, the value of postponing spatial compression until after the
full-spatial decoder prefix; \textbf{RQ4}, the contributions of temporal
and spatial components; and \textbf{RQ5}, the effects of video length
and temporal retention budget.

\begin{table}[t]
\centering
\caption{\textbf{Qwen3 VL results with input sampling at 1 FPS.} Higher accuracy (\%) and lower TTFT (ms) indicate better performance.}
\label{tab:contexttom_main}
\begingroup
\fontsize{8}{8.5}\selectfont
\setlength{\tabcolsep}{0.6pt}
\setlength{\aboverulesep}{0.7pt}
\setlength{\belowrulesep}{0.7pt}
\renewcommand{\arraystretch}{1.0}
\begin{tabular*}{\linewidth}{@{\extracolsep{\fill}}l*{14}{c}@{}}
\toprule
 & \multicolumn{8}{c}{\textbf{Video-MME}} & \multicolumn{2}{c}{\textbf{EgoSchema}} & \multicolumn{2}{c}{\textbf{LongVideo}} & \multicolumn{2}{c}{\textbf{MVBench}} \\
\cmidrule(lr){2-9}
 & \multicolumn{2}{c}{Short} & \multicolumn{2}{c}{Medium} & \multicolumn{2}{c}{Long} & \multicolumn{2}{c}{All} & \multicolumn{2}{c}{} & \multicolumn{2}{c}{\textbf{Bench}} & \multicolumn{2}{c}{(RGB)} \\
\cmidrule(lr){2-3}\cmidrule(lr){4-5}\cmidrule(lr){6-7}\cmidrule(lr){8-9}\cmidrule(lr){10-11}\cmidrule(lr){12-13}\cmidrule(lr){14-15}
\textbf{Method} & \makebox[2.3em]{Acc.$\uparrow$} & \makebox[3.1em]{TTFT$\downarrow$} & \makebox[2.3em]{Acc.$\uparrow$} & \makebox[3.1em]{TTFT$\downarrow$} & \makebox[2.3em]{Acc.$\uparrow$} & \makebox[3.1em]{TTFT$\downarrow$} & \makebox[2.3em]{Acc.$\uparrow$} & \makebox[3.1em]{TTFT$\downarrow$} & \makebox[2.3em]{Acc.$\uparrow$} & \makebox[3.1em]{TTFT$\downarrow$} & \makebox[2.3em]{Acc.$\uparrow$} & \makebox[3.1em]{TTFT$\downarrow$} & \makebox[2.3em]{Acc.$\uparrow$} & \makebox[3.1em]{TTFT$\downarrow$} \\
\midrule
\multicolumn{15}{c}{\textbf{Qwen3-VL-8B}} \\
\midrule
Full input & 79.00 & 342 & 69.22 & 877 & 60.33 & 1062 & 69.52 & 760 & 72.20 & 716 & 62.08 & 873 & 66.58$^{\ddagger}$ & 86$^{\ddagger}$ \\
Curv. merge & \textbf{78.56} & 244 & 68.00 & 687 & \underline{58.89} & 865 & \underline{68.48} & 598 & 69.60 & 489 & \textbf{61.27} & 698 & \textbf{64.79} & 70 \\
PriorTR & \underline{78.33} & 262 & \underline{68.67} & 713 & \underline{58.89} & 899 & \textbf{68.63} & 625 & \underline{70.80} & 546 & \underline{60.86} & 730 & \underline{64.56}$^{a}$ & 78$^{a}$ \\
EarlyTom & 76.11 & \underline{183} & 66.56 & \underline{473} & \textbf{59.22} & \underline{597} & 67.30 & \underline{418} & 69.20 & \underline{369} & 58.69 & 641 & 62.11 & \underline{58} \\
FrameFusion & 75.89 & 355 & 67.11 & 815 & 58.22 & 967 & 67.07 & 712 & 69.80 & 699 & 60.14 & 821 & 63.03$^{b}$ & 149$^{b}$ \\
VidCom$^{2}$ & 74.33 & 197 & 65.67 & 581 & 56.56 & 772 & 65.52 & 517 & 68.40 & 387 & 58.07 & \underline{457} & 60.26$^{b}$ & 68$^{b}$ \\
FastVID & 73.56 & 217 & 65.00 & 681 & 55.11 & 935 & 64.56 & 611 & 69.40 & 419 & 58.42 & 513 & 63.21$^{b}$ & 75$^{b}$ \\
HoliTom & 73.11 & 257 & 62.56 & 1746 & 55.00 & 3047 & 63.56 & 1683 & 70.00 & 554 & 57.25 & 656 & 61.18$^{b}$ & 74$^{b}$ \\
AIM & 73.56 & 386 & 63.00 & 1828 & 56.56 & 2917 & 64.37 & 1711 & 69.00 & 1063 & 59.13 & 1204 & 60.45$^{b}$ & 103$^{b}$ \\
\addlinespace[1pt]
\textbf{ContextTom} & 76.22 & \textbf{123.5} & \textbf{68.78} & \textbf{305.9} & 58.44 & \textbf{376.8} & 67.81 & \textbf{268.7} & \textbf{72.20} & \textbf{253.6} & 59.24 & \textbf{157.5} & 59.71 & \textbf{40.2} \\
\midrule
\multicolumn{15}{c}{\textbf{Qwen3-VL-2B}} \\
\midrule
Full input & 73.89 & 175 & 60.44 & 489 & 48.00 & 609 & 60.78 & 424 & 64.20 & 351 & 54.45 & 528 & 60.84 & 57 \\
Curv. merge & \underline{72.22} & 139 & \underline{59.44} & 421 & \underline{47.89} & 570 & \underline{59.85} & 376 & \textbf{63.60} & 279 & 53.56 & 487 & \underline{59.26} & 50 \\
PriorTR & \textbf{73.56} & 149 & \textbf{60.11} & 428 & 47.67 & 555 & \textbf{60.44} & 377 & \underline{63.20} & 296 & 53.89 & 504 & \textbf{61.58}$^{a}$ & 52$^{a}$ \\
EarlyTom & 69.67 & \underline{95} & 56.56 & \underline{259} & 47.56 & 338 & 57.93 & 231 & 59.00 & \underline{184} & 58.69 & 641 & 57.53 & \underline{40} \\
FrameFusion & 71.11 & 279 & 58.33 & 536 & 47.78 & 669 & 59.07 & 495 & 62.20 & 446 & 54.47 & 559 & 57.26$^{b}$ & 126$^{b}$ \\
VidCom$^{2}$ & 71.22 & 133 & 59.22 & 405 & 48.00 & \underline{532} & 59.48 & \underline{357} & 62.40 & 247 & 53.68 & \underline{298} & 56.76$^{b}$ & 55$^{b}$ \\
FastVID & 70.78 & 148 & 57.56 & 460 & 46.22 & 635 & 58.19 & 414 & 62.00 & 271 & \underline{56.26} & 354 & 57.13$^{b}$ & 62$^{b}$ \\
HoliTom & 67.33 & 206 & 56.22 & 1594 & 46.67 & 2975 & 56.74 & 1592 & 61.80 & 457 & 55.12 & 583 & 56.84$^{b}$ & 65$^{b}$ \\
AIM & 67.22 & 217 & 57.11 & 899 & 46.89 & 1374 & 57.07 & 830 & 61.80 & 516 & \textbf{56.43} & 622 & 57.21$^{b}$ & 79$^{b}$ \\
\addlinespace[1pt]
\textbf{ContextTom} & 70.56 & \textbf{62.4} & 58.78 & \textbf{151.0} & \textbf{48.22} & \textbf{186.8} & 59.19 & \textbf{133.4} & 61.40 & \textbf{117.8} & 52.58 & \textbf{77.1} & 55.61 & \textbf{31.3} \\
\bottomrule
\end{tabular*}
\par\vspace{3pt}
\begin{minipage}{\linewidth}
\fontsize{8}{8.5}\selectfont
\textit{Notes.} \textbf{Bold} and \underline{underlined} values indicate the best and second-best results in each column within each backbone, excluding Full input; ties share the same marking.
-- denotes an unavailable or unverified result.
MVBench: $^{a}$ 3600 examples; $^{b}$ Phase 2, 3800 examples; other listed records use 3800.
$^{\ddagger}$ is recorded as 66.58\%/86 ms, but its full-input status remains unverified in the source notes.
Budgets, dates, and evaluation sets may differ. All covers 2700 Video-MME questions.
\end{minipage}
\endgroup
\end{table}

\begin{table}[t]
\centering
\caption{\textbf{InternVL3.5 and LLaVA OneVision results with input sampling fixed at 32 frames.} Higher accuracy (\%) and lower TTFT (ms) indicate better performance.}
\label{tab:contexttom_fixedframes}
\begingroup
\fontsize{8}{8.5}\selectfont
\setlength{\tabcolsep}{0.6pt}
\setlength{\aboverulesep}{0.7pt}
\setlength{\belowrulesep}{0.7pt}
\renewcommand{\arraystretch}{1.0}
\begin{tabular*}{\linewidth}{@{\extracolsep{\fill}}l*{14}{c}@{}}
\toprule
 & \multicolumn{8}{c}{\textbf{Video-MME}} & \multicolumn{2}{c}{\textbf{EgoSchema}} & \multicolumn{2}{c}{\textbf{LongVideo}} & \multicolumn{2}{c}{\textbf{MVBench}} \\
\cmidrule(lr){2-9}
 & \multicolumn{2}{c}{Short} & \multicolumn{2}{c}{Medium} & \multicolumn{2}{c}{Long} & \multicolumn{2}{c}{All} & \multicolumn{2}{c}{} & \multicolumn{2}{c}{\textbf{Bench}} & \multicolumn{2}{c}{(RGB)} \\
\cmidrule(lr){2-3}\cmidrule(lr){4-5}\cmidrule(lr){6-7}\cmidrule(lr){8-9}\cmidrule(lr){10-11}\cmidrule(lr){12-13}\cmidrule(lr){14-15}
\textbf{Method} & \makebox[2.3em]{Acc.$\uparrow$} & \makebox[3.1em]{TTFT$\downarrow$} & \makebox[2.3em]{Acc.$\uparrow$} & \makebox[3.1em]{TTFT$\downarrow$} & \makebox[2.3em]{Acc.$\uparrow$} & \makebox[3.1em]{TTFT$\downarrow$} & \makebox[2.3em]{Acc.$\uparrow$} & \makebox[3.1em]{TTFT$\downarrow$} & \makebox[2.3em]{Acc.$\uparrow$} & \makebox[3.1em]{TTFT$\downarrow$} & \makebox[2.3em]{Acc.$\uparrow$} & \makebox[3.1em]{TTFT$\downarrow$} & \makebox[2.3em]{Acc.$\uparrow$} & \makebox[3.1em]{TTFT$\downarrow$} \\
\midrule
\multicolumn{15}{c}{\textbf{InternVL3.5-8B}} \\
\midrule
Full input & 77.67 & 367 & 62.56 & 367 & 51.67 & 367 & 63.96 & 367 & 69.80 & 369 & 60.58 & 369 & 73.34 & 365 \\
Curv. merge & 75.78 & 246 & 62.22 & 249 & 52.11 & 251 & 63.37 & 249 & \textbf{70.40} & 243 & 60.47 & 248 & 66.84 & 235 \\
PriorTR & \textbf{77.67} & 291 & \underline{63.11} & 296 & 51.11 & 298 & \underline{63.96} & 295 & \underline{69.80} & 302 & \textbf{60.63} & 298 & \underline{69.68} & 288 \\
EarlyTom & 74.22 & 188 & 59.78 & 188 & \underline{52.22} & 189 & 62.07 & 188 & 65.00 & 193 & 59.39 & 197 & 68.97 & 190 \\
FrameFusion & \underline{75.89} & 358 & 60.33 & 302 & 51.78 & 284 & 62.67 & 315 & 68.60 & 367 & 59.88 & 364 & 68.82 & 311 \\
VidCom$^{2}$ & 74.33 & \textbf{149} & 60.56 & \textbf{149} & 51.11 & \textbf{149} & 62.00 & \textbf{149} & 67.80 & \textbf{153} & 58.74 & \underline{149} & \textbf{70.61} & \underline{149} \\
FastVID & 72.00 & 175 & 58.33 & 187 & 49.89 & 191 & 60.07 & 184 & 68.20 & 173 & 55.32 & 192 & 67.71 & 150 \\
HoliTom & 68.89 & 203 & 56.44 & 186 & 48.44 & 183 & 57.93 & 191 & 66.20 & 190 & 53.46 & 193 & 63.18 & 215 \\
AIM & 71.22 & 422 & 60.00 & 423 & 51.00 & 426 & 60.74 & 423 & 68.60 & 448 & 55.92 & 440 & 64.66 & 417 \\
\addlinespace[1pt]
\textbf{ContextTom} & 75.56 & \underline{153.1} & \textbf{63.56} & \underline{153.6} & \textbf{53.67} & \underline{154.2} & \textbf{64.26} & \underline{153.8} & 67.20 & \underline{167.3} & \underline{60.51} & \textbf{148.1} & 67.53 & \textbf{145.2} \\

\midrule
\multicolumn{15}{c}{\textbf{LLaVA-OneVision-7B}} \\
\midrule
Full input & 71.11 & 293 & 56.00 & 295 & 48.33 & 295 & 58.48 & 294 & 64.20 & 298 & 55.20 & 301 & 58.39 & 294 \\
Curv. merge & \underline{70.11} & 232 & \underline{56.56} & 235 & 47.00 & 239 & \underline{57.89} & 236 & \textbf{64.60} & 239 & 54.76 & 240 & \underline{58.03} & 236 \\
PriorTR & 69.78 & 291 & 56.00 & 293 & 47.33 & 298 & 57.70 & 294 & 63.00 & 296 & 54.27 & 291 & 57.89 & 255 \\
EarlyTom & 69.44 & \textbf{157} & 55.44 & \textbf{160} & \underline{48.67} & \textbf{161} & 57.85 & \textbf{160} & 62.80 & \textbf{163} & 54.00 & \textbf{160} & 56.08 & \underline{161} \\
FrameFusion & \textbf{70.67} & 350 & 55.44 & 341 & 47.44 & 335 & 57.85 & 342 & 63.40 & 386 & \underline{54.93} & 337 & \textbf{58.39} & 275 \\
VidCom$^{2}$ & 69.00 & 169 & 55.56 & 169 & 47.89 & 169 & 57.48 & 169 & \underline{64.20} & \underline{173} & 53.84 & \underline{169} & 56.29 & 169 \\
FastVID & 68.44 & 183 & 55.33 & 189 & 46.56 & 190 & 56.78 & 187 & \textbf{64.60} & 183 & 53.20 & 189 & 56.50 & 169 \\
HoliTom & 68.00 & 225 & 54.78 & 210 & 47.33 & 208 & 56.70 & 214 & 64.00 & 219 & 52.68 & 220 & 57.47 & 234 \\
AIM & 67.78 & 256 & 54.78 & 257 & 47.78 & 260 & 56.78 & 258 & 63.20 & 272 & 52.32 & 271 & 56.24 & 255 \\
\addlinespace[1pt]
\textbf{ContextTom} & 69.56 & \underline{168.2} & \textbf{57.33} & \underline{168.9} & \textbf{48.89} & \underline{168.8} & \textbf{58.59} & \underline{168.8} & \underline{64.20} & 180.3 & \textbf{56.25} & 178.6 & 56.34 & \textbf{135.1} \\
\bottomrule
\end{tabular*}
\par\vspace{3pt}
\begin{minipage}{\linewidth}
\fontsize{8}{8.5}\selectfont
\textit{Notes.} \textbf{Bold} and \underline{underlined} values indicate the best and second-best results in each column within each backbone, excluding Full input; ties share the same marking.
-- denotes an unavailable or unverified result. Input sampling uses 32 frames for both backbones. MVBench records use 3800 examples. Budgets, dates, and evaluation sets may differ. All covers 2700 Video MME questions.
\end{minipage}
\endgroup
\end{table}

\subsection{RQ1: Accuracy and Latency Across Video Benchmarks}
\label{sec:contexttom_rq1}

Table~\ref{tab:contexttom_main} shows that ContextTom achieves the lowest TTFT
among the compared methods in every reported setting for both Qwen backbones.
On Qwen3-VL-8B, it reaches $76.22\%$ accuracy at $123.5$~ms on
Video-MME short, giving a $2.77\times$ speedup over full input
($79.00\%$, $342$~ms).
Compared with EarlyTom, it improves accuracy by $0.11$ percentage points
and reduces TTFT by $32.5\%$ on short videos; on medium videos, the
accuracy gain increases to $2.22$ points while TTFT falls by $35.3\%$.
Thus, ContextTom's shorter inference path is accompanied by higher
accuracy on both subsets.
On EgoSchema, ContextTom matches full-input accuracy at $72.20\%$
with a $2.82\times$ speedup. On LongVideoBench, it achieves
$59.24\%$ accuracy at $157.5$~ms, a $5.54\times$ speedup
from $873$~ms with a $2.84$-point accuracy decrease.
The smaller Qwen3-VL-2B also benefits: its reported TTFT speedups reach
$3.24\times$ on Video-MME medium and $6.85\times$ on LongVideoBench.

\paragraph{Fixed-frame backbones.}
Table~\ref{tab:contexttom_fixedframes} extends the evaluation to
InternVL3.5-8B and LLaVA-OneVision-7B with 32 input frames.
On Video-MME overall, ContextTom exceeds full-input accuracy by
$0.30$ and $0.11$ percentage points while delivering $2.39\times$
and $1.74\times$ TTFT speedups, respectively.
It also achieves the highest accuracy among compressed methods
on the medium, long, and overall subsets for both backbones.
TTFT remains nearly constant across duration subsets, so these
accuracy gains accompany a stable inference cost under fixed-frame sampling.
On LongVideoBench, ContextTom attains the lowest TTFT with InternVL
and the highest accuracy among compressed methods with LLaVA.
These complementary results extend the accuracy--latency benefits
beyond Qwen and variable-length 1 FPS inputs.

\subsection{RQ2: Sources of Efficiency Gains}
\label{sec:contexttom_rq2}

Development profiling in Table~\ref{tab:contexttom_efficiency} decomposes the
cost of temporal-only processing and the complete ContextTom execution path.
Decoder prefill accounts for the largest absolute saving ($90.56$~ms),
followed by vision encoding ($45.84$~ms).
Together, these stages contribute $99.1\%$ of the measured TTFT reduction.
The gains therefore span both major compute stages: temporal anchoring
reduces vision work, and a shorter late-decoder sequence reduces prefill work.
The prefix used for selection is reused by the subsequent decoder traversal.
The corresponding development budgets and additional configurations are
preserved in Table~\ref{tab:contexttom_configurations}
in Appendix~\ref{sec:development-configurations}.

\begin{table}[t]
\centering
\caption{Development-run efficiency on 900 Video-MME short questions with
Qwen3-VL-8B on one H100. Latencies are means in milliseconds; memory is
peak allocated GPU memory in the evaluator's reported GB units.
The $36.76$~ms prefix-and-selection diagnostic is included in decoder prefill.}
\label{tab:contexttom_efficiency}
\small
\setlength{\tabcolsep}{5pt}
\renewcommand{\arraystretch}{1.08}
\begin{tabular}{@{}p{0.51\linewidth}rr@{}}
\toprule
Metric & Temporal-only & ContextTom \\
\midrule
Accuracy (\%)                         & 76.00   & 76.89   \\
TTFT (ms)                             & 610.14  & 472.52  \\
Vision encoding (ms)                  & 148.24  & 102.40  \\
Pre-LLM selection and assembly (ms)    & 16.15   & 14.93   \\
Decoder prefill (ms)                  & 445.74  & 355.18  \\
Preprocessing (ms)                    & 748.57  & 750.49  \\
End-to-end latency (ms)               & 1381.38 & 1243.63 \\
Peak allocated memory (GB)            & 20.18   & 19.96   \\
\bottomrule
\end{tabular}
\end{table}

The development comparison reduces TTFT by $22.56\%$ and end-to-end latency
by $9.97\%$. Preprocessing stays near $750$~ms, explaining the smaller
end-to-end reduction: shortening the model path does not remove the
separate video-preparation cost. Peak allocated memory falls from
$20.18$ to $19.96$~GB.

\subsection{RQ3: Value of Early Spatial Context}
\label{sec:contexttom_rq3}

Table~\ref{tab:contexttom_timing} compares spatial compression before decoder
entry with ContextTom's selection after the first three decoder blocks.
Delaying selection improves accuracy by $0.60$ percentage points for
an additional $6.1$~ms TTFT. The end-to-end increase is $2.58\%$,
and peak GPU memory increases by $0.35$~GB.
This accuracy gain is obtained while only three of the 36 decoder blocks
process complete anchor grids.

\begin{table}[t]
\centering
\caption{Spatial-compression timing on Video-MME short with Qwen3-VL-8B.
End-to-end latency includes preprocessing and answer generation.
Bold indicates the highest accuracy and lowest cost.}
\label{tab:contexttom_timing}
\small
\setlength{\tabcolsep}{4pt}
\renewcommand{\arraystretch}{1.08}
\begin{tabular*}{\linewidth}{@{\extracolsep{\fill}}lrrrr@{}}
\toprule
Compression boundary & \shortstack{Acc.\ (\%)$\uparrow$}
& \shortstack{TTFT\\(ms)$\downarrow$}
& \shortstack{End-to-end\\(ms)$\downarrow$}
& \shortstack{Peak memory\\(GB)$\downarrow$} \\
\midrule
Before decoder entry & 75.62 & \textbf{117.4} & \textbf{1483.2} & \textbf{19.63} \\
After three blocks (ContextTom) & \textbf{76.22} & 123.5 & 1521.5 & 19.98 \\
\bottomrule
\end{tabular*}
\end{table}

The benefit of the prefix extends beyond the cells that survive selection.
Question states can integrate evidence from locations that are subsequently
removed, while the selected visual states carry their completed DeepStack
additions. Continuing from these states preserves the early computation
instead of rebuilding a prompt from a raw subset of visual embeddings.
The schedule separates \emph{access to evidence} from
\emph{continued computation on every location}: all anchor cells participate
in the first three blocks, but only selected representatives traverse the
remaining 33. In contrast, retaining all spatial cells throughout the decoder
(the temporal-only variant in Table~\ref{tab:contexttom_ablation}) is
$7.2$~ms slower and $0.44$ points less accurate than ContextTom.
Together, the comparisons support an intact early spatial context followed
by selective deeper processing, rather than carrying every cell through
the entire decoder.

\subsection{RQ4: Contributions of Temporal and Spatial Processing}
\label{sec:contexttom_rq4}

Table~\ref{tab:contexttom_ablation} examines five components using the
Table~\ref{tab:contexttom_main} configuration as the common reference.

\begin{table}[t]
\centering
\caption{Component ablations on Video-MME short with Qwen3-VL-8B.
$\Delta$Acc. is the change from full ContextTom in percentage points.
Bold indicates the highest accuracy and lowest TTFT.}
\label{tab:contexttom_ablation}
\small
\setlength{\tabcolsep}{6pt}
\renewcommand{\arraystretch}{1.08}
\begin{tabular}{@{}lrrr@{}}
\toprule
Variant & Acc. (\%) $\uparrow$ & $\Delta$Acc. & TTFT (ms) $\downarrow$ \\
\midrule
ContextTom (full)              & \textbf{76.22} & --       & 123.5 \\
\midrule
Uniform temporal anchors      & 75.32          & $-0.90$  & 128.0 \\
No local supplementation      & 74.56          & $-1.66$  & \textbf{120.4} \\
\midrule
Temporal compression only     & 75.78          & $-0.44$  & 130.7 \\
No query guidance             & 73.78          & $-2.44$  & 121.9 \\
Final-feature similarity only & 74.44          & $-1.78$  & 122.2 \\
\bottomrule
\end{tabular}
\end{table}

\paragraph{Temporal construction.}
Uniform selection retains the same $K$ temporal units and leaves the
subsequent stages unchanged, but reduces accuracy by $0.90$ points.
Removing local supplementation ($\eta=0$) costs $1.66$ points while saving
$3.1$~ms. Thus, the content of the temporal representatives matters beyond
their number: adaptive anchors capture representative observations, and
bounded supplementation enriches them without adding grids to the vision
transformer.

\paragraph{Query guidance and multi-level similarity.}
Removing query information from pair scoring and representative selection
causes the largest accuracy decrease ($2.44$ points) for a $1.6$~ms saving.
This identifies question relevance as a particularly valuable selection
signal. Using only final visual features for similarity costs $1.78$ points
even though all three DeepStack injections remain active.
Intermediate features therefore contribute not just through injection,
but also through the redundancy decision: the minimum similarity across
levels selects pairs that agree throughout the visual hierarchy.

\paragraph{Spatial reduction.}
With $r_s=1$, temporal-only processing reaches $75.78\%$ at $130.7$~ms.
ContextTom gains $0.44$ accuracy points and lowers TTFT by $5.51\%$.
Combining query relevance with multi-level agreement selects a smaller
continuing sequence without sacrificing the early spatial access
examined in RQ3.

\subsection{RQ5: Video Length and Temporal Retention Budget}
\label{sec:contexttom_rq5}

\paragraph{Video length.}
On the short, medium, and long Video-MME subsets,
Table~\ref{tab:contexttom_main} gives Qwen3-VL-8B TTFT speedups of
$2.77\times$, $2.87\times$, and $2.82\times$.
Absolute savings grow from approximately $219$ to $571$ and $685$~ms,
while accuracy decreases are $2.78$, $0.44$, and $1.89$ points.
Thus, the relative acceleration stays near $2.8\times$ as the longer-video
subsets offer greater absolute savings. With 1 FPS inputs, the schedule
reduces work both before vision encoding and in the deeper decoder,
addressing the larger visual workloads associated with longer videos.

\begin{table}[t]
\centering
\caption{Temporal retention on Video-MME short with Qwen3-VL-8B.
Only $r_t$ changes; $r_s=0.85$, $f=0.80$, and $p=3$ are fixed.
Retained frames and input tokens are reported means; input tokens include
visual, textual, and special tokens. Bold marks the best accuracy or TTFT.}
\label{tab:contexttom_temporal_budget}
\small
\setlength{\tabcolsep}{5pt}
\renewcommand{\arraystretch}{1.08}
\begin{tabular*}{\linewidth}{@{\extracolsep{\fill}}lrrrr@{}}
\toprule
$r_t$ & Retained frames & Input tokens & Acc.\ (\%)$\uparrow$ & TTFT (ms)$\downarrow$ \\
\midrule
0.40 & 32 & 2754.8 & 75.54 & \textbf{116.8} \\
0.52 (default) & 41 & 2988.6 & 76.22 & 123.5 \\
0.65 & 52 & 3462.2 & \textbf{76.47} & 133.2 \\
\bottomrule
\end{tabular*}
\end{table}

\paragraph{Temporal budget.}
Table~\ref{tab:contexttom_temporal_budget} reveals diminishing accuracy
returns from increased temporal retention. Raising $r_t$ from $0.40$ to
$0.52$ gains $0.68$ accuracy points for $6.7$~ms; the next increase to
$0.65$ gains $0.25$ points for $9.7$~ms.
The default therefore captures $73\%$ of the accuracy improvement across
the tested sweep while incurring $41\%$ of its additional TTFT.
Compared with $r_t=0.65$, it processes $13.7\%$ fewer multimodal input
tokens and lowers TTFT by $7.3\%$, with a $0.25$-point accuracy difference.
The first budget increase restores more useful temporal evidence per unit
of latency; further retention adds more sequence processing for a smaller
answer-quality gain. This balance motivates $r_t=0.52$ as the default,
while $0.40$ and $0.65$ provide faster and higher-accuracy alternatives.
Appendix~\ref{sec:budget-metrics} details the count definitions and
marginal-gain calculations.

\section{Conclusion}
\label{sec:conclusion}

ContextTom organizes efficient video inference around when evidence is
selected and how its computed states are reused.
The compression-timing comparison supports preserving complete spatial
access during a short decoder prefix, and the temporal-budget analysis
shows that additional observations offer diminishing accuracy returns.
Together with the component ablations, these findings favor a stage-specific
allocation of computation: representative temporal inputs, an intact
full-spatial prefix, and query-guided, multi-level selection for later layers.
This allocation of computation achieves the lowest TTFT among the compared
methods across the reported Qwen settings.
The resulting design principle is to remove temporal repetition before
vision encoding, allow early multimodal processing to use the retained
spatial evidence, and carry selected states directly into a shorter decoder
suffix. Compression then reuses the computation that makes selection
informative, connecting evidence preservation with efficient execution
in pretrained video VLMs.

\clearpage
\section*{AI Use Statement}
We used ChatGPT (OpenAI) to assist with manuscript organization,
language polishing, and LaTeX formatting. It also assisted with
interpreting the author-provided experimental results and formulating
the frame-wise score-calibration property and its proof.
All reported experimental measurements were supplied by the authors.
The authors take full responsibility for the final manuscript,
including its scientific claims, mathematical arguments, and
experimental results.

\bibliographystyle{iclr2027_conference}
\bibliography{iclr2027_conference}

\clearpage
\appendix
\hypertarget{temporal-construction}{%
\section{Temporal construction}\label{temporal-construction}}

\hypertarget{descriptors-segmentation-and-anchors}{%
\subsection{Descriptors, segmentation, and anchors}\label{descriptors-segmentation-and-anchors}}

Let \(P_{t,c}\) contain the normalized processor pixels belonging to cell \(c\) of temporal unit \(t\), including its \(m^2\) patches and temporal samples. We compute per-cell RGB means \(\mu_{t,c}\), reshape them to a spatial map, and adaptively pool to at most \(8\times8\) cells. Flattening gives \(x_t\in\mathbb R^{D_x}\), with \(d(x,y)=\|x-y\|_1/D_x\). Only these compact descriptors are transferred to the CPU for segmentation and anchor selection.

Greedy splitting recomputes the candidates of the two new intervals after every cut. Equal cut scores within an interval favor its midpoint; equal best scores across intervals favor longer intervals. Near-tied medoids favor smaller adjacent descriptor change within the interval. Anchors are emitted in temporal order. Exact medoid search costs \(O(D_x\sum_j|I_j|^2)\); distances are streamed to avoid a dense descriptor-distance tensor. Greedy splitting also incurs interval rescans.

\hypertarget{detail-and-novelty-weights}{%
\subsection{Detail and novelty weights}\label{detail-and-novelty-weights}}

Donors for anchor \(a_j\) are restricted to \(\mathcal D_j=\{u\in I_j:0<|u-a_j|\le R\}\). Let \(v_{t,c}\) be the square root of mean RGB-channel variance within \(P_{t,c}\), and let \(\delta_{u,c}=\|\mu_{u,c}-\mu_{a_j,c}\|_1/3\). The appearance scale is

\[
\sigma_j=\max\left\{\epsilon,\,
\frac{\sum_{u\in\mathcal D_j,c}\delta_{u,c}}
{\max(1,\sum_{u\in\mathcal D_j,c}\mathbf1[\delta_{u,c}>\epsilon])}\right\}.
\tag{A1}
\]

\Needspace{6\baselineskip}
With \(\epsilon=10^{-6}\), donor-cell scores are

\[
\nu_{u,c}=\delta_{u,c}/\sigma_j,\qquad
g_{u,c}=\max\left\{0,\frac{v_{u,c}-v_{a_j,c}}
{v_{u,c}+v_{a_j,c}+\epsilon}\right\},\qquad
e_{u,c}=g_{u,c}\nu_{u,c}\exp(-\nu_{u,c}).
\tag{A2}
\]

The relative detail term favors increased within-cell variation. The novelty term suppresses both duplicates and large appearance mismatches. For each cell, \(u^*(c)\) maximizes \(e_{u,c}\). Among these best donors, the top \(\lfloor\rho S\rfloor\) cells receive \(\beta_{j,c}=\eta e_{u^*(c),c}\) in Eq. (2). Empty donor sets contribute zero. Since \(g\le1\) and \(\nu\exp(-\nu)\le1/e\), the anchor weight is at least \(1-\eta/e\). With \(\eta=0.25\), every supplemented cell retains at least about \(90.8\%\) anchor weight. All \(m^2\) patches of a cell share the donor and coefficient.

The local same-cell rule requires no learned motion estimator. Because detail and novelty are measured from pixel statistics, motion, noise, and occlusion may influence donor selection.

\hypertarget{prefix-computation-and-query-scoring}{%
\section{Prefix computation and query scoring}\label{prefix-computation-and-query-scoring}}

\hypertarget{deepstack-boundary}{%
\subsection{DeepStack boundary}\label{deepstack-boundary}}

For zero-indexed decoder block \(\ell\), the prefix follows

\[
\widehat H^{(\ell+1)}=\mathcal L_\ell(H^{(\ell)};\Pi),\qquad
H^{(\ell+1)}=\widehat H^{(\ell+1)}+\mathcal I_\ell,
\qquad 0\le\ell<p,
\tag{B1}
\]

where \(\mathcal I_\ell\) inserts the corresponding DeepStack feature at visual positions for \(\ell<J\), and is zero otherwise. We use \(J\le p<L\). With \(J=p=3\), the final injection enters the visual states used by block-\(p\) Q/K scoring, before those states undergo another complete decoder block.

The sequence remains causal: visual positions precede the question and do not attend to it. Question states have access to visual evidence, and the selection scores are query-dependent. This distinction determines which states carry question information at the compression boundary.

\hypertarget{causally-normalized-importance}{%
\subsection{Causally normalized importance}\label{causally-normalized-importance}}

Let \(\mathcal Q\) be the text suffix strictly after the last video-end marker, and \(\mathcal V\) the visual positions. We apply the pretrained block-\(p\) normalization, Q/K projections, Q/K head normalization, and RoPE. For head \(h\),

\[
A^{(h)}_{q,i}=
\frac{\exp(\langle Q^{(h)}_q,K^{(h)}_i\rangle/\sqrt{d_h})}
{\sum_{u\le q}\exp(\langle Q^{(h)}_q,K^{(h)}_u\rangle/\sqrt{d_h})},
\quad i\le q,
\qquad
s_i=\frac{\bar A_i}{\sum_{v\in\mathcal V}\bar A_v},\quad
\bar A_i=\frac{\sum_{q\in\mathcal Q,h}A^{(h)}_{q,i}}{|\mathcal Q|n_h}.
\tag{B2}
\]

All causally visible text and visual keys enter the denominator. Query rows exclude timestamps and empty wrappers before the final video-end marker. Q is computed only for the suffix; K covers the full prefix. Score products use FP32. The scoring operation adds Q/K work at block \(p\), whose projections are evaluated again when that block processes the reduced sequence. Its score-product cost is approximately \(O(n_h|\mathcal Q|n_0d_h)\).

\hypertarget{spatial-selection-and-state-continuation}{%
\section{Spatial selection and state continuation}\label{spatial-selection-and-state-continuation}}

\hypertarget{pair-construction-and-deterministic-selection}{%
\subsection{Pair construction and deterministic selection}\label{pair-construction-and-deterministic-selection}}

Horizontal pairs are used if \(W\ge H\) and \(W>1\); otherwise pairs are vertical. They are disjoint, stay within each frame, and do not wrap across rows or columns. An unpaired boundary cell survives. Similarities in Eq. (3) use the final and DeepStack visual features, before decoder processing.

Query ranks are stable ascending ranks across all \(KS\) visual cells, normalized to \([0,1]\). After selecting the top \(C\) pairs per frame and global top \(B\), each chosen pair retains the member with greater \(s_i\). Equal importance favors the larger squared norm of the final visual embedding. The evaluated configuration uses zero spatial donor weight, so the selected state is gathered without averaging. It applies no additional absolute-similarity veto. Integer rounding and candidate capacities can yield a higher retention than the nominal \(r_s\).

\hypertarget{positions-and-caches}{%
\subsection{Positions and caches}\label{positions-and-caches}}

Prompt assembly keeps original text, timestamps, and modality wrappers, including wrappers of removed temporal units. Position construction advances through the original frame scaffold. The spatial gather retains existing position IDs rather than renumbering surviving cells.

After prefill, layer \(\ell\) stores \(n_0\) KV entries if \(\ell<p\), and \(n_1\) entries otherwise. Each generated token is appended to every layer\textquotesingle s own history. New text positions start at \(1+\max(\Pi)\) and increase monotonically. This implementation supports a single unpadded sequence with SDPA.

\hypertarget{complete-inference-algorithm}{%
\subsection{Complete inference algorithm}\label{complete-inference-algorithm}}

\begin{Verbatim}[fontsize=\footnotesize]
Input: processed video, prompt, frozen VLM, rt, rs, f, p
1. Form RGB descriptors; split T temporal units into K intervals.
2. Choose medoid anchors and score nearby same-cell donors.
3. Embed patches; supplement selected cells before vision block 0.
4. Encode K full grids; collect final and DeepStack features.
5. Assemble the prompt and original multimodal position IDs.
6. Run blocks 0,...,p-1 once with the complete anchor grids.
7. Score post-video text queries at the input to block p.
8. Rank local pairs by multi-level similarity and query importance.
9. Select B pairs under per-frame caps; retain one cell per pair.
10. Gather surviving H^(p) states and positions, keeping all text.
11. Continue through blocks p,...,L-1 to obtain first-token logits.
12. Generate with full early-layer and reduced later-layer caches.
Output: generated answer
\end{Verbatim}

\hypertarget{execution-and-measurement}{%
\section{Execution and measurement}\label{execution-and-measurement}}

\hypertarget{cost-and-implementation}{%
\subsection{Cost and implementation}\label{cost-and-implementation}}

For decoder-layer cost \(C_{\mathrm{dec}}(n)\) at length \(n\), the computation decomposes approximately as

\[
\begin{aligned}
C_{\mathrm{total}}\approx{}&
C_{\mathrm{embed}}(T)+C_{\mathrm{vision}}(K)
+pC_{\mathrm{dec}}(n_0)+(L-p)C_{\mathrm{dec}}(n_1)\\
&+C_{\mathrm{temporal}}+C_{\mathrm{score}}+C_{\mathrm{select}}.
\end{aligned}
\tag{D1}
\]

The decoder has components linear in sequence length and attention interactions quadratic in length. The early prefix is paid once. The practical implementation batches equal-length independent vision-frame attention into one SDPA call per block, uses fused RMSNorm and rotary-position kernels, and assembles only retained visual features. Vision attention remains within temporal units. The main configuration uses the dense temporal-fusion backend and greedy segmentation.

\Needspace{21\baselineskip}
\hypertarget{development-configuration-and-timing}{%
\subsection{Development configuration and timing}\label{development-configuration-and-timing}}

\begin{longtable}[]{@{}
  >{\raggedright\arraybackslash}p{(\columnwidth - 2\tabcolsep) * \real{0.5000}}
  >{\raggedright\arraybackslash}p{(\columnwidth - 2\tabcolsep) * \real{0.5000}}@{}}
\caption{Development profiling configuration and recorded reference results.}
\label{tab:development_settings}\\
\toprule\noalign{}
\begin{minipage}[b]{\linewidth}\raggedright
Setting
\end{minipage} & \begin{minipage}[b]{\linewidth}\raggedright
Value
\end{minipage} \\
\midrule\noalign{}
\endhead
\bottomrule\noalign{}
\endlastfoot
Backbone & Frozen Qwen3-VL-8B-Instruct \\
Hardware and execution & One H100; single request; eager SDPA \\
Data & Video-MME short; 300 videos, 900 questions; no subtitles \\
Video preprocessing & 1 FPS; long edge 512 pixels \\
Temporal configuration & \(r_t=0.52\), \(\lambda_d=1\), \(R=2\), \(\rho=\eta=0.25\) \\
Spatial configuration & \(p=3\), \(r_s=0.85\), \(f=0.80\), \(\lambda_q=0.7\) \\
Development-run result & 692/900 correct; 76.89\% accuracy; 472.52 ms mean model-side TTFT \\
Temporal-only development reference & 684/900 correct; 76.00\%; 610.14 ms; \(r_t=0.50\) \\
\end{longtable}

Model-side TTFT starts immediately before temporal selection and visual preparation and ends when first-token logits are available, with synchronized timing. It includes temporal selection, vision encoding, sequence assembly, the decoder prefix, spatial scoring and selection, and later decoder blocks. Video decoding and processor/tokenizer preprocessing precede this interval.

Each question receives an untimed same-input warmup, followed by a fresh timed forward. Hidden states and caches are recomputed. The spatial-selection diagnostic includes the reused prefix and is already part of decoder prefill.

Mean TTFT comprises 102.40 ms of vision computation, 14.93 ms of pre-LLM selection/assembly, and 355.18 ms of decoder prefill including spatial selection. Mean preprocessing is 750.49 ms. Recorded end-to-end latency is 1243.63 ms, excluding the untimed warmup. The 22.56\% latency reduction compares complete configurations, including temporal-budget and execution differences, on the development subset used to select the configuration.

\subsection{Development configurations}
\label{sec:development-configurations}

Table~\ref{tab:contexttom_configurations} records the development configurations
used alongside the stage-level profiling in
Table~\ref{tab:contexttom_efficiency}.
The temporal and spatial retention ratios and execution paths vary across
these development runs. Pre-decoder local-pair compression reduces TTFT
relative to temporal-only processing, and query-guided two-pass compression
recovers part of its accuracy at a smaller spatial budget.
The complete development configuration reaches $76.89\%$ accuracy at
$472.52$~ms, improving over its temporal-only reference by $0.89$ accuracy
points and $22.56\%$ lower TTFT.

\begin{table}[htbp]
\centering
\caption{Development configurations on Video-MME short using Qwen3-VL-8B
and one H100. Temporal and spatial columns give nominal retention ratios;
spatial retention is relative to the temporally compressed stream.
Configurations include different budgets and execution paths.}
\label{tab:contexttom_configurations}
\small
\setlength{\tabcolsep}{4pt}
\renewcommand{\arraystretch}{1.08}
\begin{tabular}{@{}p{0.40\linewidth}rrrr@{}}
\toprule
Configuration & Temporal & Spatial & Acc. (\%) & TTFT (ms) \\
\midrule
Temporal-only                     & 0.50 & 1.00 & 76.00 & 610.14 \\
Pre-decoder local-pair compression & 0.50 & 0.80 & 74.33 & 534.78 \\
Query-guided two-pass compression  & 0.50 & 0.65 & 75.11 & 480.81 \\
ContextTom                        & 0.52 & 0.85 & \textbf{76.89} & \textbf{472.52} \\
\bottomrule
\end{tabular}
\end{table}

\subsection{Development token flow}
\label{sec:development-token-flow}

The development trace contains a mean of $5734.77$ visual tokens before
compression and $2988.16$ after temporal anchoring.
All anchor cells traverse the first three decoder blocks and receive the
three DeepStack injections. Spatial selection retains $2540.32$ visual
states for the final 33 blocks, or $44.30\%$ of the original visual tokens.
The spatial stage retains $85.01\%$ of anchor cells in aggregate, matching
the nominal $r_s=0.85$ up to integer rounding and candidate capacities.
These are visual-only counts from the development profile;
Table~\ref{tab:contexttom_temporal_budget} reports multimodal input counts
from the temporal-budget experiments.

\subsection{Compression-timing and budget metrics}
\label{sec:budget-metrics}

\paragraph{Compression timing.}
The two configurations in Table~\ref{tab:contexttom_timing} apply spatial
reduction before decoder entry or after the first three decoder blocks.
The delayed configuration is the full ContextTom reference in
Table~\ref{tab:contexttom_main}.
End-to-end latency includes preprocessing and answer generation;
peak GPU memory is the reported maximum during inference.
The $6.1$~ms TTFT difference is the difference between the complete measured
paths. Reusing the prefix lets its computation contribute both to selection
and to the final answer, while later layers use the selected sequence.

\paragraph{Count definitions.}
Retained frames denotes the mean number of video frames preserved after
temporal compression and passed to subsequent visual processing.
Input tokens denotes the mean length of the compressed multimodal input,
including visual tokens, textual prompt tokens, and special tokens:
\[
N_{\mathrm{input}}=N_{\mathrm{visual}}+N_{\mathrm{text}}+N_{\mathrm{special}}.
\]
The frame count measures temporal retention, while the input-token count
measures the reported multimodal sequence size. The decoder itself uses
the two lengths $n_0$ and $n_1$ defined in
Section~\ref{overview-compress-time-early-preserve-space-through-the-prefix},
with full anchor grids in the prefix and spatially selected states afterward.

\paragraph{Marginal accuracy returns.}
For adjacent settings $a$ and $b$, define
\[
G_{a\rightarrow b}=
\frac{\operatorname{Acc}(b)-\operatorname{Acc}(a)}
{\operatorname{TTFT}(b)-\operatorname{TTFT}(a)}.
\]
Using percentage-point accuracy differences, the measured gains are
$G_{0.40\rightarrow0.52}=0.68/6.7\approx0.1015$ points/ms and
$G_{0.52\rightarrow0.65}=0.25/9.7\approx0.0258$ points/ms.
The earlier budget increment therefore yields about $3.9$ times the
accuracy gain per added millisecond.
Relative to $r_t=0.40$, the default realizes
$0.68/(76.47-75.54)=73.1\%$ of the sweep's accuracy improvement for
$6.7/(133.2-116.8)=40.9\%$ of its latency increase.
Relative to $r_t=0.65$, it uses
$(3462.2-2988.6)/3462.2=13.7\%$ fewer input tokens and reduces TTFT by
$(133.2-123.5)/133.2=7.3\%$.
These calculations describe the measured temporal-retention settings and
explain the operating-point choice in RQ5.

\end{document}